# Random-Walk Metaball-Imaging Discrete Element Lattice Boltzmann Method for 3D Solute Transport in Fluid-Particle Systems with Complex Granular Morphologies

Yifeng Zhao[a,b,c,d], Pei Zhang[b,c], Stan Z. Li[d], S.A. Galindo-Torres[b,c,*]

[a]*College of Environmental and Resources Science, Zhejiang University, 866 Yuhangtang Road, Hangzhou, 310058, Zhejiang Province, China*
[b]*Key Laboratory of Coastal Environment and Resources of Zhejiang Province, School of Engineering, Westlake University,, 18 Shilongshan Road, Hangzhou, 310024, Zhejiang Province, China*
[c]*M3 lab, School of Engineering, Westlake University, 18 Shilongshan Road, Hangzhou, 310024, Zhejiang Province, China*
[d]*AI lab, School of Engineering, Westlake University, 18 Shilongshan Road, Hangzhou, 310024, Zhejiang Province, China*

**Abstract**

Solute transport in fluid-particle systems is a fundamental process in numerous scientific and engineering disciplines. The simulation of it necessitates the consideration of solid particles with intricate shapes and sizes. To address this challenge, this study proposes the Random-Walk Metaball-Imaging Discrete Element Lattice Boltzmann Method (RW-MI-DELBM). In this model, we reconstruct particle geometries with the Metaball-Imaging algorithm, capture the particle behavior using the Discrete Element Method (DEM), simulate fluid behavior by the Lattice Boltzmann Method (LBM), and represent solute behavior through the Random Walk Method (RWM). Through the integration of these techniques with specially designed boundary conditions, we achieve to simulate the solute transport in fluid-particle systems comprising complex particle morphologies. Thorough validations, including analytical soluutions and experiments, are performed to assess the robustness and accuracy of this framework. The results demonstrate that the proposed framework can accurately capture the complex dynamics of solute transport under strict mass conservation. In particular, an investigation is carried out to assess the influence of particle morphologies on solute transport in a 3D oscillator, with a focus on identifying correlations between shape features and dispersion coefficients. Notably, all selected shape features exhibited strong correlations with the dispersion coefficient, indicating the significant influence of particle shapes on transport phenomena. However, due to the complexity of the relationship and the limited number of simulations, no clear patterns could be observed. Further comprehensive analyses incorporating a broader range of shape features and varying conditions are necessary to fully comprehend their collective influence on the dispersion coefficient. The proposed RW-MI-DELBM offers a promising framework to study solute transport in fluid-particle systems with complex morphologies, ensuring strict mass conservation of the solute. It has the potential to be used in various applications, including the study of pollutant dispersion in environmental systems, optimization of heat transfer in industrial processes, and enhancement of drug delivery in pharmaceutical applications.

---

*Principal Corresponding author
*Email address:* s.torres@westlake.edu.cn (S.A. Galindo-Torres)

## 1. Introduction

Solute transport involves the transfer of dissolved substances, such as concentration or pollutants, through a medium. This transport can be driven by various mechanisms, including advection and diffusion.[1, 2, 3]. It is a fundamental concept in various fields of science and engineering and plays a crucial role in understanding and analyzing the behavior of many complex systems[4, 5]. In most real-world scenarios, the solute transport involves not only fluid but also particles with intricate geometries and sizes, ranging from irregular granular materials to complex biological structures. The presence of such non-spherical particle significantly influences the transport phenomena, altering flow patterns, concentration distributions, and the overall system behavior, which adds further complexity to the study of solute transport[6, 7, 8]. The dispersion of pollutants in rivers is a good example. Sediment particles can exhibit diverse shapes, ranging from irregular aggregates to elongated fibers, which significantly influence the transport characteristics of pollutants through intricate inter-particle interactions[9]. Furthermore, in the field of drug delivery, the shape of human cells assumes paramount importance as it directly impacts the diffusion and distribution of drug solute within the human body. And this can significantly influence the therapeutic effectiveness and overall treatment outcomes, making it imperative to meticulously select appropriate drug particle shapes to optimize drug delivery strategies[10, 11]. Similarly, particle shape plays a crucial role in solute transport phenomena in chemical engineering processes. Consider the case of fluidized bed reactors, where the shape distribution of catalyst particles profoundly affects various aspects such as mixing efficiency, minimum fluidization velocity, and segregation behaviors[12, 13, 14]. Therefore, understanding the intricate interplay between solute transport and solid particles of different shapes in fluid systems is pivotal for comprehending a broad spectrum of phenomena and developing effective strategies across domains such as environmental science, geophysics, fluid dynamics, and biomedical engineering.

However, the intricate nature of this system poses significant obstacles for analytical methods, which often struggle to provide accurate results due to the involved complexities in particle shape. Consequently, numerical simulations have emerged as an indispensable tool for unraveling the complex dynamics of this solute-particle-fluid system. In general, those different quantities are tackled with different methods and coupled together to predict the overall behavior of the solute transport system[15, 16, 17].

For the solid particle, such as sediments, the Discrete Element Method (DEM) is frequently utilized. DEM treats particles as rigid entities and considers their motions based on the Newton-Euler equations. Many successful applications have been made on the modelling of realistic non-spherical particles with DEM through X-ray Computed Tomography (XRCT). The main differences between those methods are the shape reconstruction methods and their corresponding contact frameworks. For instance, Spherical-Harmonic (SH) DEM applies the SH function to reconstruct non-spherical particle shape and interparticle overlapping to do contact detection[18, 19, 20]. Level-set DEM utilizes the level-set function to handle the shape capture and a look-up mechanism for contact detection[21, 22]. Poly-superellipsoid DEM utilizes an assembly of eight separate superellipsoids to capture particle morphologies and tackle contact detection with an efficient hybrid Levenberg-Marquardt (LM) and Gilbert-Johnson-Keerthi (GJK) algorithms[23, 24]. Another approach, signed distance field (SDF) DEM, defines particle shapes using the SDF function and incorporates an energy-conserving contact theory for contact detection[25]. Recently, the authors employ the Metaball function to reconstruct realistic particle morphologies[26] and couple it with DEM in a gradient-based method [27, 28, 29]. This Metaball DEM approach strikes a balance between particle shape representation and computational efficiency, making it suitable for studying fluid-particle systems with large amout of complex-shaped particles[29]. In this paper, the Metaball DEM approach will be used to explore the dependence of solute transport on grain morphology.

For the fluid, the Lattice Boltzmann Method (LBM) is usually implemented[30] with DEM as a DELBM scheme to solve the fluid-particle system. The utilization of LBM in such coupling is driven by its distinct advantages. Firstly, LBM exhibits excellent parallelization efficiency owing to the localized nature of the collision operator, which is particularly advantageous in computationally demanding fluid-particle simulations. Additionally, the inherent kinetic nature of LBM empowers it to effectively address intricate moving boundary conditions through straightforward algorithms. On this basis, the DELBM coupling schemes can be categorized into two main approaches: diffuse interface and sharp interface. The diffuse interface schemes handle the discontinuity at solid-fluid boundaries by smoothing the interface. Example can be found like the Immersed Boundary Methods (IBM), where the influence of solid boundaries is replaced by a smoothed external force field. Another notable approach is the partially saturated cells method (PSM), which incorporates solid boundaries through the representation of solid volume fraction. Those methods are widely used in DEM-LBM coupling due to their widespread application and simplicity. However, their non-physical representation limits their accuracy[31]. For instance, IBM can only achieve first-order accuracy in simulating porous media flows[32], and PSM systematically underestimates permeability[28]. On the contrary, the sharp interface schemes treat solid boundaries without smoothing. It follows the bounce-back rule, where fluid molecules contacting the solid surface are reflected back to the fluid domain with opposite velocity. The accuracy of it is further improved in [33, 34, 35], achieving second-order accuracy in space. In our previous work, we also introduce an efficient sharp interface scheme for DELBM, where the contact detection efficiency is significantly enhanced by utilizing information from LBM[36].

For the solute particle, two main approaches can be identified: grid-based methods and particle-based methods. Grid-based methods involve the allocation and updating of solute quantities on a grid. A notable example is the work by Derksen[16], who employ a finite-volume approach within a LBM simulation coupled with an event-driven hard-sphere DEM to solve solute transport. Another intriguing study is conducted by Xia et al.[37], where a high-order immersed boundary method (IBM) based on ghost-cells is implemented. This method offers computational efficiency advantages and can ensure consistent treatment of various boundary conditions. Additionally, Wang et al.[38] adopt a double distribution function (DDF) approach to address heat transfer in fluid-particle systems. Moreover, Lei et al. [39] develop an enhanced approach for predicting heat transfer in gas-solid fluidized beds by incorporating improvements in the filtered interphase heat transfer coefficient (IHTC) closure. Lastly, Xu et al. [17] utilize a DDF-based LBM method to simulate heat transfer in fluid-particle systems, successfully identifying the influence of heat on the settling behavior of elliptical particles. However, despite their advantages, grid-based methods suffer from certain limitations, particularly when dealing with moving boundaries. One prominent issue is the contradiction of solute mass conservation. This is caused by the interpolation procedure in many no-gradient boundary conditions and the switches between solid and fluid nodes[15]. Such contradiction can lead to inaccuracies and inconsistencies in the simulation results, especially in applications where the need to ensure proper solute mass conservation is a crucial aspect. In contrast, particle-based methods offer an alternative approach where the solute quantity is represented by discrete, independent particles, and its transport is simulated through the collective movement of these particles, considering advection and diffusion effects. This methodology avoids the mass conservation issue associated with grid-based methods by directly tracking the individual particles and their interactions, rather than relying on grid manipulation. By simulating the transport of solute quantities based on the movements and interactions of these discrete particles, particle-based methods offer a more accurate representation of solute mass conservation. In our recent research [15], we have presented a noteworthy example of the particle-based approach. In this framework, the transport of solute is simulated with particles using the random particle method. And the obtained results demonstrate excellent agreement with analytical solutions and experimental observations, thereby confirming the effectiveness of particle-based methods in accurately capturing solute transport phenomena. However, it is important to acknowledge that the framework proposed in that research is limited to 2D conditions and fluid-particle systems comprising only spherical particles. To address more general scenarios and encompass a wider range of applications, there is still a clear need for a more comprehensive approach ca-pable of simulating solute transport in 3D fluid-particle systems that involve complex-shaped particles and adhere to the solute mass conservation.

In this paper, we propose a Random-Walk Metaball-Imaging Discrete Element Method (RW-MI-DELBM), designed specifically for simulating 3D solute transport in fluid-particle systems with complex granular morphologies. This method couples several techniques to achieve its objectives. Firstly, the Metaball-function and the accompanying Metaball-Imaging algorithm are employed to accurately reconstruct the intricate particle shapes, including realistic irregular geometries. On this basis, DEM is applied to tackle the particle behavior. And LBM is utilized to simulate the fluid behavior. Finally, the Random Walk Method (RWM), a particle-based technique, is employed to represent the transport of the solute quantity. The proposed method has undergone rigorous validation through a series of analytical and experimental cases, confirming its validity and effectiveness. Notably, a showcase involving a thermostatic shaking water bath exemplifies the substantial impact of particle shape on the dispersion coefficient of the solute.

The remainder of the paper is organized as follows. Section 2 provides a comprehensive overview of the technical details and methodology employed in this study. Section 3 presents a series of validation cases, encompassing both analytical and experimental scenarios, confirming the accuracy and reliability of the proposed method. In Section 4, an application case study involving a thermostatic shaking water bath is presented, highlighting the influence of particle shape on the dispersion process of the solute. Finally, Section 5 summarizes the major contributions of this work and provides concluding remarks.

## 2. Methodology

In this methodology, several critical model assumptions are established. The initial assumption posits that the solute particle adheres to Fick's law[40], signifying that no reactions occur within the system. Furthermore, the model assumes that the shape of solid particles is non-spherical and exhibits complex and realistic geometries, rather than being perfectly spherical. In terms of the fluid, the model characterizes it as a Newtonian flow, ensuring compliance with the non-slip condition at the fluid-particle interface[15].

These assumptions collectively facilitate a focused investigation of the solute-particle-fluid system, particularly under the influence of particle motions. Such framework is representative of, and applicable to, a diverse range of physical processes commonly encountered in various scientific and engineering contexts.

*2.1. Random Walk Method for the solute*

The random walk method (RWM) is a widely-used numerical technique for simulating the advection-diffusion process in transport phenomena[40], which is governed by the advection-diffusion equation:

$$\frac{\partial C}{\partial t} + \boldsymbol{u} \cdot \nabla C = D\Delta C \tag{1}$$

where $C$ denotes the concentration, $t$ represents time, $\boldsymbol{u}$ signifies the fluid velocity field, and $D$ corresponds to the molecular diffusion coefficient.

RWM models the aforementioned process by tracking the movement of individual solute particles, which is driven by the Brownian motion resulting from the continuous interaction with surrounding fluid molecules[40, 41]. The method incorporates information concerning the solute's position based on the local velocity from the fluid (advection) and a random displacement from the diffusion coefficient (diffusion):

$$\boldsymbol{x}_r(t+\delta t) = \boldsymbol{x}_r(t) + \boldsymbol{u}\left(\boldsymbol{x}_r, t\right)\delta t + \boldsymbol{z}\sqrt{2D\delta t} \tag{2}$$

where $\boldsymbol{x}_r$ denotes the position of the solute particle. $\boldsymbol{u}$ is the local fluid velocity at the particle's location. And $\boldsymbol{z}$ is a random vector drawn from a Gaussian distribution with zero mean and unit variance. The term $\boldsymbol{u}\left(\boldsymbol{x}_r, t\right)\delta t$ characterizes the deterministic movement of particles due to fluid flow, while $\boldsymbol{z}\sqrt{2D\delta t}$, accounts for the random, Brownian motion of particles induced by molecular diffusion. By combining these two components, the RWM can effectively simulate the overall particle movement in the advection-diffusion process with strict mass conservation[15].

*2.2. Metaball-Imaging Discrete Element method for the particle*

The Metaball-Imaging Discrete Element method (MI-DEM) is previously introduced by the authors[27] as an effective means to simulate and investigate the mechanical properties of realistic granular materials.

Within this framework, the Metaball function is employed to represent particles:

$$f(\boldsymbol{x}) = \sum_{i=1}^{n} \frac{\hat{k}_i}{\left(\boldsymbol{x}-\hat{\boldsymbol{x}}_i\right)^2} \tag{3}$$

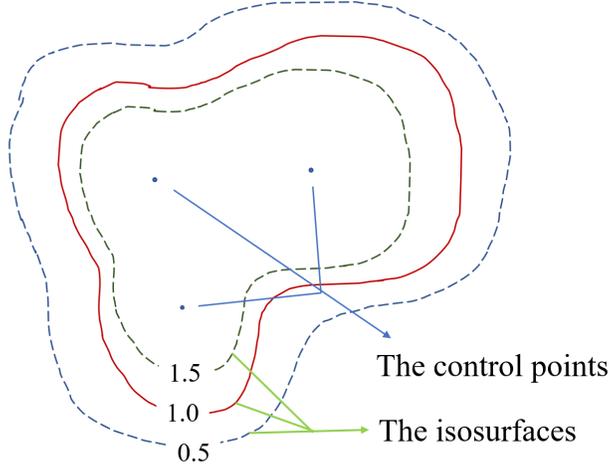

Figure 1: A sketch map of the Metaball function.

where $n$ denotes the number of control points. $\hat{\boldsymbol{x}}_{\boldsymbol{i}}$ refers to the position of the $i$-th control point, which serves as a skeleton for the parameterized shape. And $\hat{k}_i$ represents the positive coefficient determining the weight of the $i$-th control point. As an implicit function, it defines shapes with a series of isosurfaces, which consist of points in the space satisfying certain function values, around some control points (See Fig. 1). Building on this foundation, the Metaball-Imaging algorithm is developed to reconstruct complex particle morphologies. This process consists of three steps: data preprocessing, capturing the principal outer contour using sphere-clustering, and refining the extracted contour through gradient-search techniques. Due to page constraints, further details can be found in [26, 29].

DEM is a popular numerical technique designed to simulate the motion and behavior of particulate systems, such as granular materials, powders, and aggregates[42]. In DEM, each particle is treated as a discrete entity, where the translational motion is governed by Newton's equations and the rotational motion is described using the conservation of angular momentum model:

$$\begin{cases} m_i \boldsymbol{a}_i = m_i \boldsymbol{g} + \sum_{j=1}^{N} \boldsymbol{F}_{ij}^c + \boldsymbol{F}_i^e \\ \frac{d}{dt}\left(\mathbf{I}_i \boldsymbol{\omega}_i\right) = \sum_{j=1}^{N} \boldsymbol{T}_{ij}^c + \boldsymbol{T}_i^e \end{cases} \quad (4)$$

where $m_i$ and $\boldsymbol{a_i}$ represent the mass and acceleration of the $i$-th particle, respectively. $\boldsymbol{g}$ denotes the gravitational acceleration. $\boldsymbol{F}_{ij}^c$ signifies the contact force between the particle $i$ and its neighboring particle $j$, while $\boldsymbol{F}_i^e$ corresponds to the external force. The inertia tensor is given by $\mathbf{I}_i$, and $\boldsymbol{\omega}i$ indicates the angular velocity. Finally, $\boldsymbol{T}_{ij}^c$ and $\boldsymbol{T}_i^e$ represent the torques arising from contact forces and external forces, respectively.

Here, the contact force $\boldsymbol{F}^c$ is described by the linear spring-dashpot model[43]. The normal component of $\boldsymbol{F}^c$ is given by: $F_n^c = k_n\delta + \eta_n (\boldsymbol{v_j} - \boldsymbol{v_i}) \cdot \boldsymbol{n}$, where $k_n$ represents the normal spring stiffness. $\delta$ denotes the particle overlapping distance. $\eta_n$ refers to the normal damping coefficient. $\boldsymbol{n}$ stands for the unit normal vector. And $\boldsymbol{v_i}$ and $\boldsymbol{v_j}$ correspond to the velocities of the $i$-th and its neighbor $j$-th particles, respectively. The tangential component of $\boldsymbol{F}^c$ is determined as:

$$\begin{cases} F_t^c &= \min\left(\mu_s F_n^c, F_{t0}^c\right) \\ F_{t0}^c &= \left\|-k_t \boldsymbol{\xi} - \eta_t (\boldsymbol{v_j} - \boldsymbol{v_i}) \cdot \boldsymbol{t}\right\| \end{cases} \quad (5)$$

where $\mu_s$ denotes the static friction coefficient. $k_t$ represents the tangential spring stiffness. $\eta_t$ refers to the damping coefficient. $\boldsymbol{\xi}$ is the tangential spring, and $\boldsymbol{t}$ is the unit tangential vector.

Using Metaball function for reconstructing Discrete Element Method (DEM) particles, collisions between Metaballs and between Metaballs and walls can be accurately handled using gradient-based methods. This requires the determination of three parameters: the contact point $\boldsymbol{x}_{cp}$, the contact direction $\boldsymbol{n}$, and the overlap $\delta$. To avoid unwanted intersection issues, the collision is first transformed into a sphero-Metaball based contact [27, 29], in which the original Metaball is approximated by a combination of a zoomed internal Metaball and a dilated sphere with radius $R_s$. Subsequently, an optimization problem is introduced to locate the closest points on the studied Metaball set or between Metaball and wall. This optimization problem is defined as:

$$\begin{aligned} \text{Minimize} \quad & f_0(\boldsymbol{x}) + f_1(\boldsymbol{x}) \\ \text{Subject to} \quad & c_{tol} < |f_0(\boldsymbol{x})| < 1, \quad c_{tol} < f_1(\boldsymbol{x}) < 1 \end{aligned} \quad (6)$$

where $f_0(\boldsymbol{x})$, $f_1(\boldsymbol{x})$ represent functions of two Metaballs; $c_{tol}$ is the tolerance to avoid $f_0(\boldsymbol{x}) + f_1(\boldsymbol{x}) = 0$ when $\|\boldsymbol{x}\| \to \infty$. Under this constraint, the obtained solution will satisfy the condition that the gradient of optimization target, Eq. 6 equals 0:

$$\nabla(f_0(\boldsymbol{x}) + f_1(\boldsymbol{x})) = \mathbf{0} \quad (7)$$

In collisions between Metaballs, the Newton-Raphson method is employed to find the local minimum point $\boldsymbol{x_m}$. The closest points on the Metaballs

($\boldsymbol{x_{c0}}$ and $\boldsymbol{x_{c1}}$) can be approximated by:

$$\begin{cases} \boldsymbol{x}_{c0} = \boldsymbol{x}_m + q_0 \nabla f_0(\boldsymbol{x}_m) \\ \boldsymbol{x}_{c1} = \boldsymbol{x}_m + q_1 \nabla f_1(\boldsymbol{x}_m) \end{cases} \quad (8)$$

By combining Taylor series expansion and imposing $f_0(\boldsymbol{x_{c0}}) = 1$, $q_0$ and $q_1$ can be explicitly expressed to calculate $\boldsymbol{x_{c0}}$ and $\boldsymbol{x_{c1}}$. Consequently, the required three parameters for the collision between Metaballs can be determined as:

$$\begin{cases} \boldsymbol{x}_{cp} = \boldsymbol{x}_{c0} + (R_{s0} - 0.5\delta)\boldsymbol{n} \\ \boldsymbol{n} = \frac{\boldsymbol{x}_{c0} - \boldsymbol{x}_{c1}}{\|\boldsymbol{x}_{c0} - \boldsymbol{x}_{c1}\|} \\ \delta = R_{s0} + R_{s1} - \|\boldsymbol{x}_{c1} - \boldsymbol{x}_{c0}\| \end{cases} \quad (9)$$

In Metaball-wall collisions, the problem is simplified to finding a point $\boldsymbol{x_{cw}}$ on the wall with a normalized gradient of $(-1, 0, 0)$ with respect to the rotated Metaball function, corresponding to a point $\boldsymbol{x_{cm}}$ on it. This point is searched using the Newton-Raphson method. Then, the closest point on the Metaball to the wall is determined as:

$$\boldsymbol{x}_{cp} = \boldsymbol{x}_{cw} + q^W \nabla f^W(\boldsymbol{x}_{cw}) \quad (10)$$

With $f(\boldsymbol{x}_{cw}) = 1$ and Taylor series expansion, the required three parameters for the collision between Metaball and wall can be defined as:

$$\begin{cases} \boldsymbol{x}_{cp} = \boldsymbol{x}_{cw} + 0.5\delta\boldsymbol{n} \\ \boldsymbol{n} = \frac{\boldsymbol{x}_{cm} - \boldsymbol{x}_{cw}}{\|\boldsymbol{x}_{cm} - \boldsymbol{x}_{cw}\|} \\ \delta = R_s - \|\boldsymbol{x}_{cm} - \boldsymbol{x}_{cw}\| \end{cases} \quad (11)$$

2.3. Lattice Boltzmann Method for the fluid

The Lattice Boltzmann Method (LBM) is a widely used Computational Fluid Dynamics technique grounded in kinetic theory[30]. Rather than solving fluid flows at the continuum scale, LBM employs the Boltzmann equation to describe fluid motion with distribution functions as the fundamental variable. The distribution function $G_i(\boldsymbol{x}, t)$ signifies the probability of a fluid molecule occupying a lattice cell at position $\boldsymbol{x}$ with a specific discrete velocity $\boldsymbol{e}_i$ at time $t$[36]. In this framework, D3Q19 model, involving the evolution of distribution functions according to the equation provided in the previous response, is utilized[29]:

$$G_i(\boldsymbol{x} + \boldsymbol{e}_i \delta t, t + \delta t) - G_i(\boldsymbol{x}, t) = -\frac{1}{\tau} \left[ G_i(\boldsymbol{x}, t) - G_i^{eq}(\boldsymbol{x}, t) \right] \quad (12)$$

where $\tau$ is the relaxation time, $\boldsymbol{e}_i$ denotes the discrete lattice velocities, and $G_i^{eq}$ is the equilibrium distribution function. The equilibrium distribution function is defined as:

$$G_i^{eq}(\boldsymbol{x}, t) = \omega_i \rho \left[ 1 + \frac{\boldsymbol{e}_i \cdot \boldsymbol{u}}{c_s^2} + \frac{(\boldsymbol{e}_i \cdot \boldsymbol{u})^2}{2c_s^4} - \frac{\boldsymbol{u} \cdot \boldsymbol{u}}{2c_s^2} \right] \quad (13)$$

where $\omega_i$ are the lattice weights, $\rho$ represents the fluid density, $\boldsymbol{u}$ is the macroscopic fluid velocity, and $c_s$ is the speed of sound in the lattice. The fluid density and velocity are calculated from the distribution functions as follows:

$$\rho(\boldsymbol{x}, t) = \sum_{i=0}^{18} G_i(\boldsymbol{x}, t) \quad (14)$$

$$\boldsymbol{u}(\boldsymbol{x}, t) = \frac{1}{\rho(\boldsymbol{x}, t)} \sum_{i=0}^{18} G_i(\boldsymbol{x}, t) \boldsymbol{e}_i \quad (15)$$

The LBM algorithm consists of two main steps: collision and streaming[30]. During the collision step, the distribution functions relax towards their equilibrium values, while the streaming step updates the distribution functions based on their neighboring values. By simulating the evolution and collision of these distribution functions, LBM can recover the Navier-Stokes equations, allowing the study of various flow phenomena[29]

2.4. Intergration of RWM, LBM and MI-DEM

2.4.1. Coupling of RWM and LBM: Connecting Solute and Fluid Grid

To couple RWM and LBM, the solute particle linearly interpolates the velocity field from the LBM grid, which contains it. While the solute particle is still in the Brownian motion as a random walk process (Eq. 2). On this basis, the solute particle information is also transferred into the LBM grid as a solute value to save the saving cost:

$$C(\boldsymbol{x}, t) = \frac{m_s n_s(\boldsymbol{x}, t)}{\delta x^3} \quad (16)$$

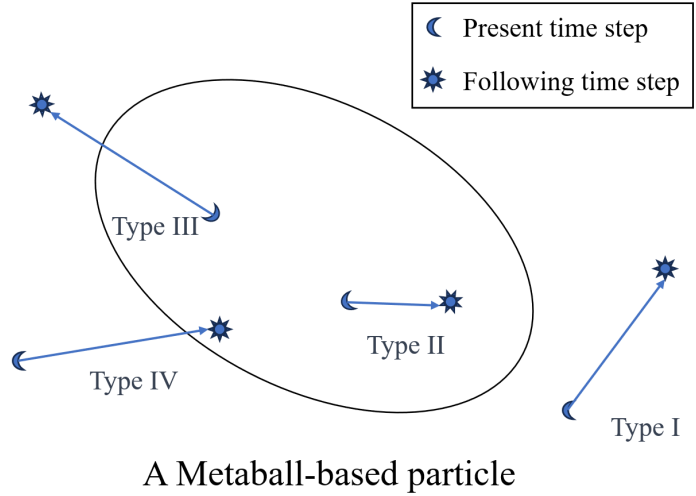

Figure 2: Four types of solute-Metaball particle relationships.

where $\boldsymbol{x}$ is the node position, $\delta x$ is the grid size, $m_s$ is the mass of individual solute particle, $\varepsilon_n$ is the solid occupation fraction and $n_s(x,t)$ is the total number of solute particles within the node at time t.

*2.4.2. Coupling of RWM and MI-DEM: Investigating Solute-Particle Dynamics*

To couple RWM and MI-DEM, the non-penetration boundary condition needs to be satisfied. The treatment of solute-Metaball particle boundaries involves three distinct types: Dirichlet, and no-flux.

Prior to the introduction of Dirichlet and no-flux boundaries, it is important to note that only four types of solute-Metaball particle relationships exist during the simulation, considering both the present and following time steps (See Fig. 2): I. outside the Metaball particles in both time steps; II. inside the Metaball particles in both time steps; III. inside the Metaball particle in the present time step and outside the Metaball particle in the following time step; IV. outside the Metaball particle in the present time step and inside the Metaball particle in the following time step.

The Dirichlet boundary primarily involves type II particles, which are inside the Metaball particles in both time steps. To address this boundary, a "refill" approach is utilized. As demonstrated in Eq. 3, the Metaball function is differentiable, allowing for the calculation of the gradient:

$$\nabla f(\boldsymbol{x}) = \sum_{i=1}^{n} \frac{-2\hat{k}_i (\boldsymbol{x} - \hat{\boldsymbol{x}}_i)}{|\boldsymbol{x} - \hat{\boldsymbol{x}}_i|^4} \qquad (17)$$

The type II solute particles are pushed out of the Metaball particle along the calculated gradient from the above function and randomly placed within a circumscribed sphere of radius $\sqrt{2D\delta t}$ on the particle surface. Due to narrow gaps in particulate flow, solute particles may undergo multiple reflections. To simplify the process, these solute particles are returned to their original location. Note that the treatment of soulte particles at the begining step is different. To avoid local concentration surge, those solute particles will be simply removed. While the above simplifications may not adhere to microscopic properties, numerous numerical validations in this paper, including author previous study [15], demonstrate its efficacy.

The no-flux boundary primarily concerns type IV particles, which are situated outside the Metaball particle in the present time step and inside it in the following time step. To handle this situation, a specular reflection is implemented. The process involves two steps: first, locating the intersection point between the solute and Metaball particle, and second, reflecting the overlapping trajectory of the solute particle within the corresponding Metaball particle. Since the Metaball particle is represented as a differentiable implicit function, the above two steps can be easily executed as following.

In step 1, the intersection point $\boldsymbol{x}_{inter}$ is located using binary search on the trajectory of the solute particle (See Algorithm 1). This is because the Metaball function values along the trajectory are continuous and in a specific order, with the intersection point having a Metaball function value $f(\boldsymbol{x}_{inter})$ equal to 1. The search range is initialized with $\boldsymbol{x}_p$ having a Metaball function value less than 1 and $\boldsymbol{x}_f$ having a Metaball function value greater than 1. During each iteration of the binary search, we compute the midpoint $\boldsymbol{x}_{mid}$ between the current search range $\boldsymbol{x}_p$ and $\boldsymbol{x}_f$, and evaluate the Metaball function at $\boldsymbol{x}_{mid}$, denoted as $f(\boldsymbol{x}_{mid})$. If $f(\boldsymbol{x}_{mid})$ is less than 1, the search range is updated to be between $\boldsymbol{x}_{mid}$ and $\boldsymbol{x}_f$. Otherwise, the search range is updated to be between $\boldsymbol{x}_p$ and $\boldsymbol{x}_{mid}$. The binary search continues until a point on the trajectory with a Metaball function value close enough to 1 is found, with a convergence criterion of $1.0 \times 10^{-3}$ or a maximum of 200 iterations. If no suitable point is found, an error is reported. Once a suitable intersection point is found, it can be considered as the point where the solute particle intersects the Metaball particle. This intersection point can then be used in the subsequent step.

---
**Algorithm 1** Binary Search for Solute-Particle Intersection
---
    **Input:** Two endpoints of solute particle trajectory $\boldsymbol{x}_p$, $\boldsymbol{x}_f$
    **Output:** Intersection point $\boldsymbol{x}_{inter}$
1: Set convergence criterion $\epsilon = 1.0 \times 10^{-3}$, maximum iterations $I_{max} = 200$
2: Initialize search range: $\boldsymbol{x}_p$ has $f(\boldsymbol{x}_p) < 1$ and $\boldsymbol{x}_f$ has $f(\boldsymbol{x}_f) > 1$
3: Set iteration counter $i = 1$
4: **while** $i \leq I_{max}$ **do**
5:     Compute midpoint $\boldsymbol{x}_{mid}$ between $\boldsymbol{x}_p$ and $\boldsymbol{x}_f$
6:     Compute $f(\boldsymbol{x}_{mid})$
7:     **if** $|f(\boldsymbol{x}_{mid}) - 1| < \epsilon$ **then**
8:         Set $\boldsymbol{x}_{inter} = \boldsymbol{x}_{mid}$ and break
9:     **else if** $f(\boldsymbol{x}_{mid}) < 1$ **then**
10:         Update search range to be between $\boldsymbol{x}_{mid}$ and $\boldsymbol{x}_f$
11:     **else**
12:         Update search range to be between $\boldsymbol{x}_p$ and $\boldsymbol{x}_{mid}$
13:     **end if**
14:     Increment $i$ by 1
15: **end while**
16: **if** $i > I_{max}$ **then**
17:     Report error: No suitable intersection point found within $I_{max}$ iterations
18: **end if**
19: **Return:** Intersection point $\boldsymbol{x}_{inter}$

---

In step 2, reflection is carried out using linear calculation (See Algorithm 2). The intersection point $\boldsymbol{x}_{inter}$ from the previous step separates the trajectory vector of the solute particle into the outside vector $\boldsymbol{t}_o$ and the inside vector $\boldsymbol{t}_i$, which can be expressed as:

$$\begin{cases} \boldsymbol{t}_o = \boldsymbol{x}_{inter} - \boldsymbol{x}_p \\ \boldsymbol{t}_i = \boldsymbol{x}_f - \boldsymbol{x}_{inter} \end{cases} \quad (18)$$

The gradient of the Metaball function $f$ at $\boldsymbol{x}_{inter}$ gives the normal vector $\boldsymbol{n} = f'(\boldsymbol{x}_{inter})$ at the intersection point. Based on this, the reflected vector

$t_r$ can be computed as:
$$t_r = t_i - 2(t_i \cdot n)n \tag{19}$$

---

**Algorithm 2** Reflection of Solute Particle Trajectory

    **Input:** the intersection point $x_{inter}$, the endpoints of the trajectory $x_p$ and $x_f$, and the Metaball function $f$.
    **Output:** the reflected vector $t_r$.

1: **Calculate the inside and outside vectors:**
$$\begin{cases} t_o = x_{inter} - x_p \\ t_i = x_f - x_{inter} \end{cases} \tag{20}$$

2: **Compute the normal vector:**
$$n = \nabla f(x_{inter}) \tag{21}$$

3: **Compute the reflected vector:**
$$t_r = t_i - 2(t_i \cdot n)n \tag{22}$$

4: **Return:** The reflected vector $t_r$.

---

It is important to highlight that in the simulation after initialization, the existence of type II and III solute particles can only be observed after the movement of DEM particles due to the implementation of Dirichlet and no-flux boundary conditions. Specifically, the type II solute particles are handled with the Dirichlet boundary condition. And the type III solute particles just move out of the DEM particles, without any boundary constraints. Furthermore, the type I particle is also not subjected to any boundary conditions, enabling unrestricted movement within the system.

*2.4.3. Coupling of MI-DEM and LBM: Investigating Fluid-Particle Interactions*

In the coupling of DEM and LBM, the boundary conditions, hydrodynamic forces, and refilling algorithms are carefully chosen. The interpolated bounce-back (IBB) method is used for the moving boundary condition to capture the influence of Metaball particles on the fluid and impose the non-slipping boundary condition. This scheme categorizes nodes into fluid, solid, and boundary nodes, represented by "s" (closest solid node), "w" (wall), "f" (boundary node), and "ff" (neighboring fluid node of "f"), as shown in Fig.3.

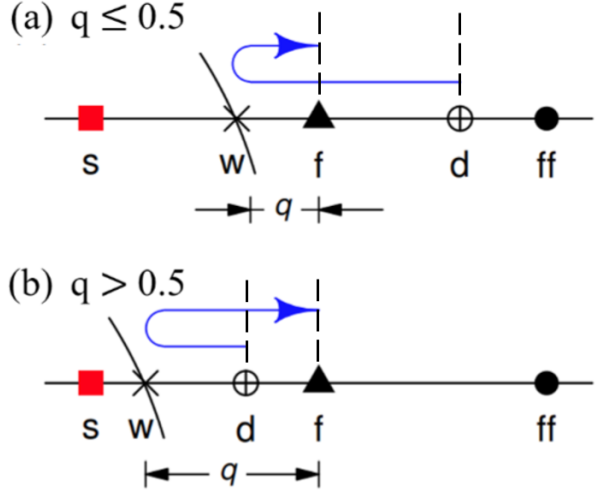

Figure 3: The schematic of the interpolated bounce-back rule, illustrating two different scenarios corresponding to various conditions of $q$.

The distribution function for a point departing from $\boldsymbol{x}_d$ with velocity $e_{i'}$ and returning to $\boldsymbol{x}_f$ with velocity $e_i$ after hitting the wall is defined as:

$$G_i(\boldsymbol{x}_f, t + \Delta t_{LBM}) = G_{i'}^+(\boldsymbol{x}_d, t) + 6\omega_{i'}\rho_f \frac{\boldsymbol{e}_i \cdot \boldsymbol{u}_w}{C^2}, \qquad (23)$$

where $\omega_i$ represents the LBM weights for each discrete direction, and $\rho_f$ denotes the fluid density. The particle surface velocity $\boldsymbol{u}_w$ is calculated using the translational and angular velocities at the $j$th particle's centroid $\boldsymbol{x}_{pj}$. The post-collision distribution function $G_{i'}^+(\boldsymbol{x}_d, t)$ is decomposed into equilibrium and non-equilibrium parts:

$$G_{i'}^+(\boldsymbol{x}_d, t) = G_{i'}^{eq}(\rho_d, \boldsymbol{u}_d) + G_{i'}^{neq}(\boldsymbol{x}_d, t), \qquad (24)$$

The velocity $\boldsymbol{u}_d$ is explicitly expressed through linear interpolation:

$$\boldsymbol{u}_d = \frac{1}{3}\boldsymbol{u}_d^* + \frac{2}{3}\boldsymbol{u}_d^{**}. \qquad (25)$$

where $\boldsymbol{u}_d^* = \begin{cases} 2q\boldsymbol{u}_f + (1-2q)\boldsymbol{u}_{ff}, & q \leqslant 0.5, \\ \frac{1-q}{q}\boldsymbol{u}_f + \frac{2q-1}{q}\boldsymbol{u}_w, & q > 0.5, \end{cases}$ , $q$ can be obtained by solv-ing $f(\boldsymbol{x}_f + q\boldsymbol{e}_i) = c_0$, and $c_0$ is a Metaball function value depending on $R_s$. As for $\boldsymbol{u}_d^{**}$, it is equal to $(\frac{1-q}{1+q}\boldsymbol{u}_{ff} + \frac{2q}{1+q}\boldsymbol{u}_w)$.

For the hydrodynamic forces, a momentum exchange method (MEM) is selected to accurately represent fluid-particle interactions. A Galilean-invariant MEM is implemented to maintain the Galilean invariance principle and effectively reduce numerical noise.

As DEM particles move through LBM cells, the missing $G_i$ from the particle must be initialized. A local refilling algorithm based on the bounce-back rule is applied, and the reinitialized distribution function is defined as:

$$G_i(\boldsymbol{x}_{new}, t) = G_{i'}(\boldsymbol{x}_{new}, t) + 6\omega_{i'}\rho_f\frac{\boldsymbol{e}_i \cdot \boldsymbol{u}_w}{C^2}, \tag{26}$$

If $G_{i'}(\boldsymbol{x}_{new}, t)$ does not exist, the equilibrium refilling is used:

$$G_i(\boldsymbol{x}_{new}, t) = G^{eq}(\rho_0, \boldsymbol{u}_{new}) \tag{27}$$

where $\boldsymbol{u}_{new} = \boldsymbol{v}_{pj} + \boldsymbol{w}_{pj} \times (\boldsymbol{x}_{new} - \boldsymbol{x}_{pj})$.

## 3. Validation

In order to verify the accuracy and reliability of the proposed framework, we have conducted a series of simulations encompassing various aspects of solute transport processes, including both analytical solutions and experimental cases. In the author's previous work[15], the effects of solute particle count in simulations, boundary treatment schemes and coupled MI-DELBM model were comprehensively examined. Building upon these findings, the following validation cases focus on the advection and diffusion of solute quantities, as well as the interactions between solute fields and larger granular materials within the 3D context.

### 3.1. Validation of Advection-diffusion in 3D Flow Conditions

Under no flow conditions, the transportation of a fixed point source follows the Fick's law:

$$\frac{\partial C}{\partial t} = D\nabla^2 C \tag{28}$$

with the following initial boundary condition:

$$C(\mathbf{x}, 0) = M\delta(\mathbf{x} - \mathbf{x_0}) \tag{29}$$

and the boundary condition at the infinity:

$$\lim_{|\mathbf{x}|\to\infty} C(\mathbf{x}, t) = 0 \tag{30}$$

where $C$ is the solute concentration, $t$ is time. $D$ is the diffusion coefficient. $M$ represents the total mass of solute particles, where $M = \sum m_s$. $\delta$ is the Dirac delta function. And $\mathbf{x_0}$ is the location of the point source. The analytical solution for the concentration distribution can be obtained using the Fourier transform method:

$$C(\mathbf{x}, t) = \frac{M}{(4\pi Dt)^{3/2}} \exp\left(-\frac{|\mathbf{x} - \mathbf{x_0}|^2}{4Dt}\right) \tag{31}$$

The validated simulation is conducted in a 0.10×0.10×0.10 m domain with a point source of 1,000,000 solute particles located at its center. And the concentration $C$ here is defined as the number of particles $n_p$ in unit space $s_u$:

$$C = \frac{n_p}{s_u} \tag{32}$$

The simulation utilized a diffusion coefficient of 1.0 ×$10^{-3}$ $m^2/s$ and assumed a constant zero flow field, indicating no fluid flow in the simulation. To approximate the analytical solution for a brief time period, a periodic boundary condition is applied to the domain. The domain and point source configuration are depicted in Fig. 4, (a). The space step and time step in simulation are 1.0 ×$10^{-3}$ $m$ and 2.0×$10^{-4}$ $s$.

The simulation results for each axis are presented in 5. It can be observed that the simulation results for all three axes match well with the analytical solution given in Eq. 31.

Under constant flow conditions, the transportation of a point source follows the advection-diffusion equation with the same boundary condition of Eq. 29:

$$\frac{\partial C}{\partial t} + \mathbf{u} \cdot \nabla C = D\nabla^2 C \tag{33}$$

Where $\mathbf{u}$ is the constant flow velocity vector. The analytical solution for the concentration distribution can be obtained as follows:

$$C(\mathbf{x}, t) = \frac{M}{(4\pi Dt)^{3/2}} \exp\left(-\frac{|\mathbf{x} - \mathbf{x_0} - \mathbf{u}t|^2}{4Dt}\right) \tag{34}$$

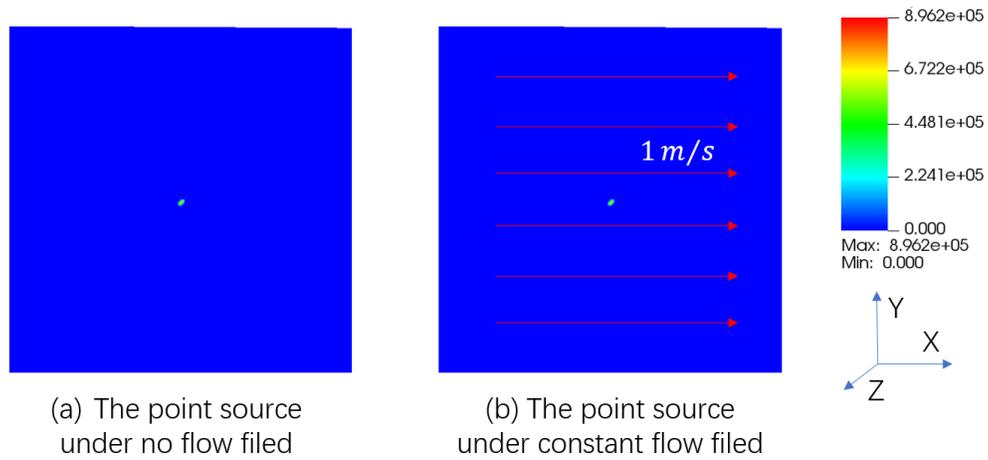

Figure 4: Sketch map of the simulation domain for point source under no flow field and constant flow filed. The color in figure stands for the number of solute particle. In (b), the red arrows indicate the direction of the flow, with "1 m/s" denoting the flow speed

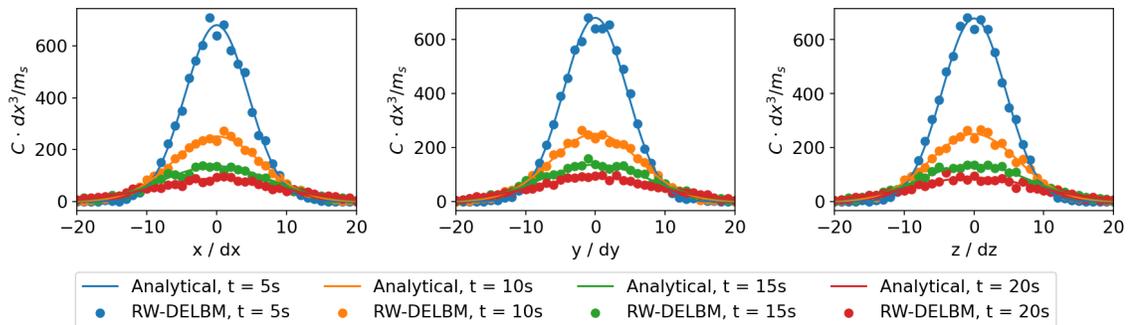

Figure 5: Concentration profile of the fixed solute point source under a constant zero flow field along the three axises

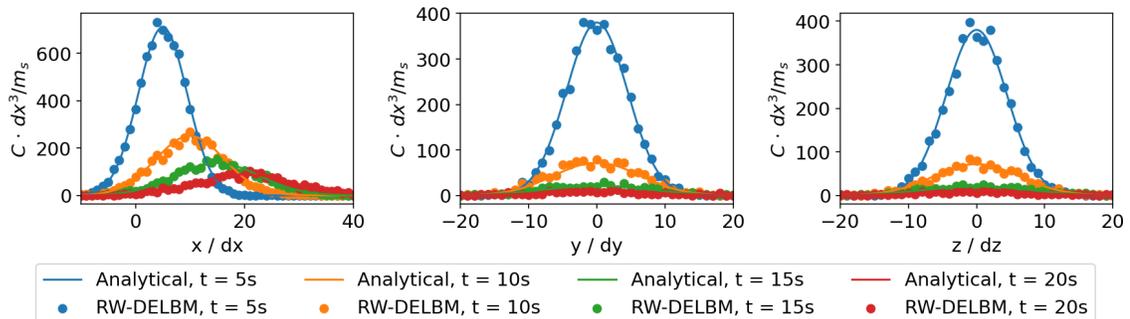

Figure 6: Concentration profile along the three axises

The corresponding validated simulation employs identical settings for the no-flow conditions, with the exception of a constant flow velocity vector $\mathbf{u} = (1, 0, 0)$ m/s, indicating a unidirectional flow across the domain. The domain and point source configuration are depicted in Fig. 4, (b).

The simulation results for each axis are presented in Fig. 6. It can be observed that the simulation results for all three axes match well with the analytical solution given in Eq. 34.

Together, these two examples prove that the proposed framework can accurately simulation the 3D advection-diffusion process in fluid fields.

## 3.2. Validation of Interactions between Solute and Solid Particles

To further validate the proposed framework with moving boundaries, especially on non-spherical particles, a series of experiments are performed. The experiments focused on irregularly shaped particles and involved observing their traversal through colorful solute concentration profiles.

The detailed experimental setup can be observed in Fig. 7.They are conducted in a rectangular acrylic container of with internal dimension: 0.04 × 0.04 × 0.150$m$, with a white background to provide contrast. A colored solute consisting of Indigo, which is denser than water (with density of 1400 $kg/m^3$), is injected slowly to the bottom of the container. Through a period of static settling, a stable boundary line is achieved (See Fig. 7, a). Three distinct irregular-shaped particles, particles A - C as illustrated in Fig. 9, are specifically designed and 3D-printed using Resinene, with a density of 1100 $kg/m$. All three particles are designed to have the same volume of a sphere with a radius of 0.01 $m$. In experiments, those irregular particles are released by tweezers in a completely submerged state. The initial release orientation is carefully chosen to align the maximum projection plane as perpendicular to the settling direction as possible. Such release mode can enable the particle to settle in a steady state. The entire process is recorded using a high-speed camera (CANON EOS 5D MARK IV), and the resulting video is processed with OpenCV [44] to extract the concentration profile of the colored water solution. It is worth noting that, although the Indigo solute is denser than water, the boundary between them is not completely flat. In fact, the solute-water boundary is a 3D surface with a complex geometry that is not easily observable. For this reason, we averaged over the boundary instead of recording the highest point of solute, as shown in Fig. 7, (b).

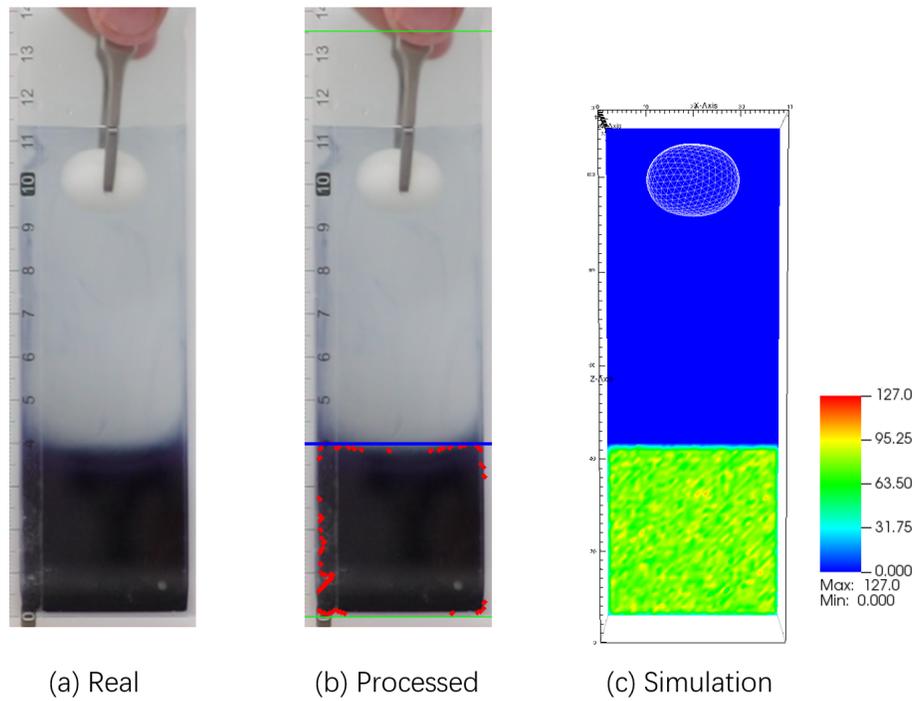

(a) Real      (b) Processed      (c) Simulation

Figure 7: Sketch map of the simulation and experiment for the validation of interactions between the solute and solid particle. (a) depicts the experimental setup, (b) is the imaged-processed one with sign on the solute-fluid interface and (c) displays the reconstructed simulation setup, where the white mesh stands for the particle and colorful bulk for the solute.

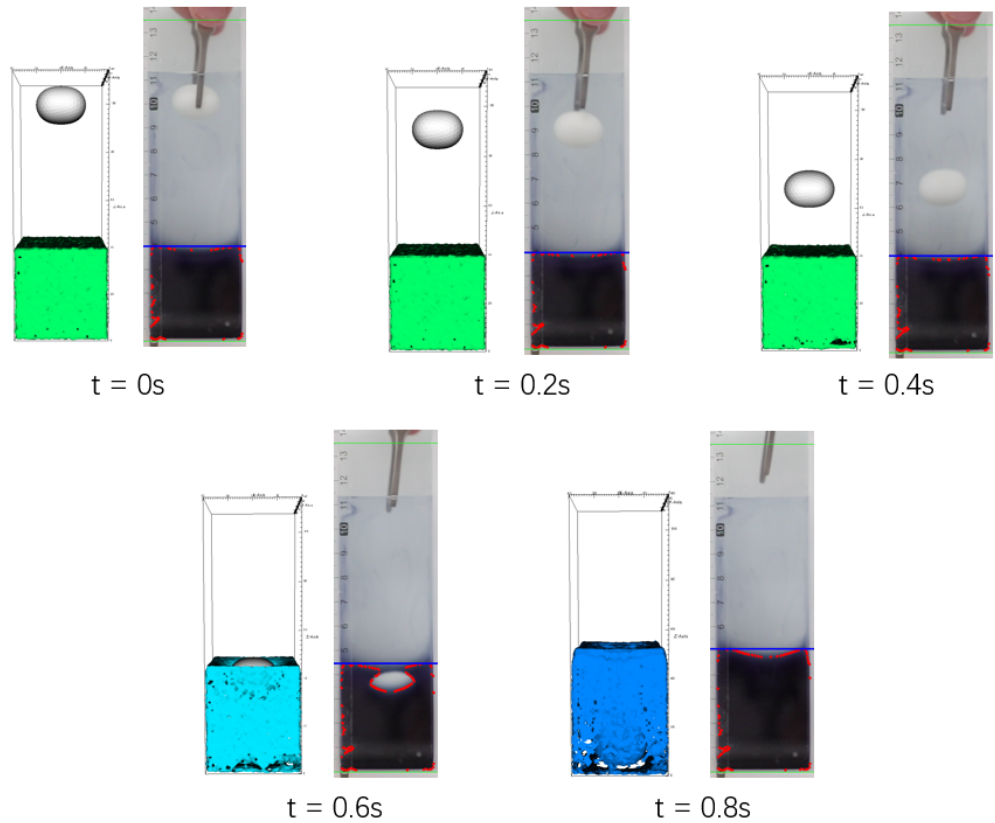

Figure 8: Time series of the simulation and the experiment for the validation of interactions of the solute and solid particle. Each time step includes a combination of snapshots from both the simulation and experiment, where the simulation snapshot is on the left, while the experiment snapshot on the right. Note that the colorful surfaces in the simulation snapshots represent the 3D isosurfaces of solute particle concentrations.

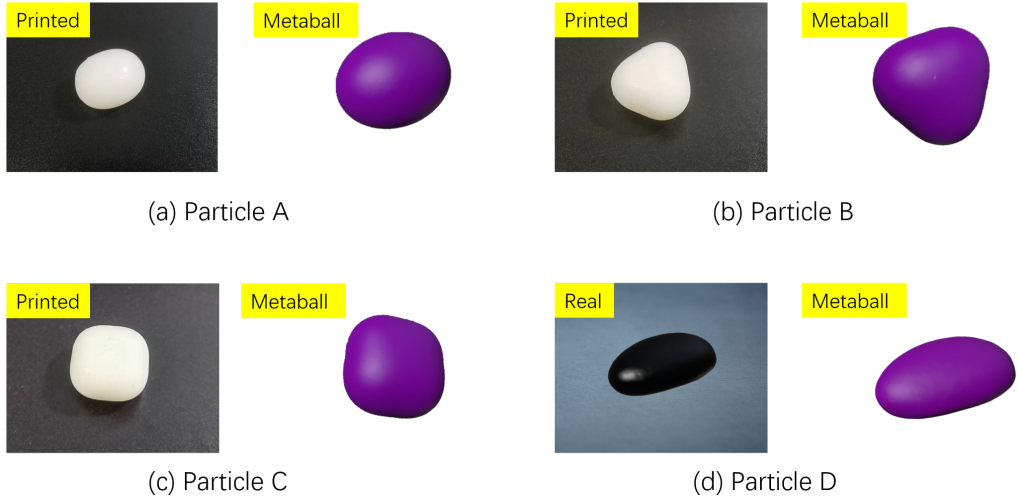

Figure 9: The morphologies of printed and Metaball-reconstructed particles. (a) represents an ellipsoid structure comprising two control points, while (b) showcases a shape constructed with three control points arranged in a triangular configuration. Furthermore, (c) displays a shape formed by four control points arranged in a rectangular arrangement. Lastly, (d) illustrates a realistic cobblestone and its Metaball model obtained through Metaball imaging techniques. Here, set of (a) to (c) is used for the validation of interactions between the solute and solid particles. And set of (a) to (d) is applied in the application of the proposed framework. Note that the purple meshes are only for visualizations. The particles in simulations are represented in the form of Metaball functions.

To make one to one comparisons, the settings of experiments and simulations are kept as identical as possible (See Fig. 7, c). The viscosity of fluid is set to be $1.0 \times 10^{-3} N \cdot s/m^2$, the viscosity value of water in $20°C$. The diffusion coefficient of the solute is selected as the 2e-9 $m^2/s$. The Metaball-reconstructed particles are shown in Fig. 9. It is worth noting that those surface meshes of particles are only for visualization and those particles are presented as Metaball functions in the simulation system. The LBM and DEM time steps are set as: $2.0\times10^{-4}$ and $2.0\times10^{-6}s$. The LBM space step is set as $1.0 \times10^{-3}m$.

In the following discussion, the Reynolds number $Re$ and the Péclet number $Pe$ are involved and defined as:

$$Re = \frac{\rho_f U_p L}{\mu} \qquad (35)$$

$$Pe = \frac{U_p L}{D} \qquad (36)$$

where $\rho_f$ represents the density of fluid, $U_p$ is the peak velocity of the fluid and $L$ stands for the longest axis length of the particle.

Since MI-DELBM scheme for fluid-particle systems implemented in the proposed framework is already verified in author's previous works[29, 27, 28]. Here, we focus on the analysis of the changes in the heights of averaged solute-fluid boundaries. Fig. 10, 11, and 12 illustrate the boundary-height and time series for particles A, B, and C, respectively, in both the experiments and simulations. Overall, the simulation results exhibit a favorable agreement with the experimental observations during the settling processes. Similar patterns are observed across all three particles. Notably, each particle undergoes two distinct stages: the approaching stage and the interaction stage.

The approaching stage denotes the beginning phase in which the particle is released and gradually approaches the solute-water boundary. During this stage, the particle-solute interaction has not yet commenced, and the dispersion of solutes is primarily governed by diffusion. Given the condition $Pe < 1$ and the relatively small value of the diffusion coefficient, minimal alterations in the boundary can be observed in both the experimental and simulation results. During the interaction stage, the particle and solute begin to interact, leading to gradual increases in the boundary heights. These changes are primarily driven by the advection through the flow field, considering $Pe > 1$. An excellent agreement can be observed during the initial period of interaction between the experimental and simulation results. However, small deviations with relative errors within 6% become noticeable in the following period. One of the major error sources could be attributed to the approximation of the initial condition of the solute-water boundary. As stated earlier, the particle-solute boundaries in the experimental setup exhibit complex 3D geometries that are challenging to capture under limited experimental conditions. In the simulations, we employed averaged boundary heights as substitutes to reconstruct the boundary. Additionally, the particles are released manually and due to the irregular morphologies of the particles, strict control over the initial releasing orientation is not feasible. Therefore, the simulation setup represents an approximation. Overall, the aforementioned approximations are reasonable, as they yield a good match during the initial period. However, errors are continuously introduced into the simulation, which can accumulate over time and result in larger discrepancies between the simulation and the observed data as the number of time steps increases.

It is worth noting that the ending pattern of particle C differs from the other particles, as the experimental result surpasses the simulation. This discrepancy can be primarily attributed to the intricate shape of particle C, as shown in Fig. 9 (c). This makes it more challenging to fix using tweezers, which leads to a greater disparity in the initial settling gesture between simulations and experiments compared to the other particles. In fact, particle C undergoes noticeable tumbling during its settling process, further intensifying the advection behavior of the solute. Besides, the experimental setting of particle C leads to larger values of *Re* and *Pe*. Consequently, the above factors together make water-solute boundary in the experimental setup exceed that of the simulation.

In conclusion, the experimental and simulation results exhibit excellent agreement, with errors falling within an acceptable range. This serves as a further evidence of the effectiveness and robustness of the proposed simulation framework.

## 4. A 3D Case Study on the Solute Dispersion in Fluid-particle Systems of complex granular morphologies with the Oscillator

Solid particle shape can have significant impacts on solute transport in a fluid-particle system, especially when considering the movement of immersed particles. It is easy to understand that the movement of irregularly shaped particles can have a different effect on the fluid flow patterns with spherical ones, leading to various variations in turbulence and velocity. These alterations, in turn, can influence the dispersion, mixing, and concentration gradients of solutes within the surrounding fluid. Moreover, the shapes of particles directly affect their surface interactions, further influencing the overall solute transport dynamics. While direct evidence may be limited[15, 45], the above factors suggest that particle shape can indeed play an important role in solute transport in fluid-particle systems. In this context, the proposed (RW-MI-DELBM) framework provides a valuable computational tool for investigating the effects of morphological parameters in those physical processes.

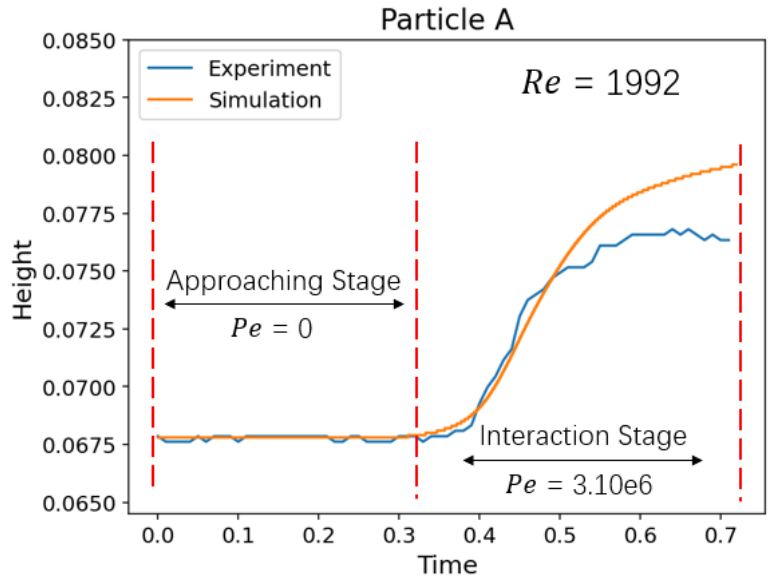

Figure 10: The comparison of the experimental and simulated results on Particle A.

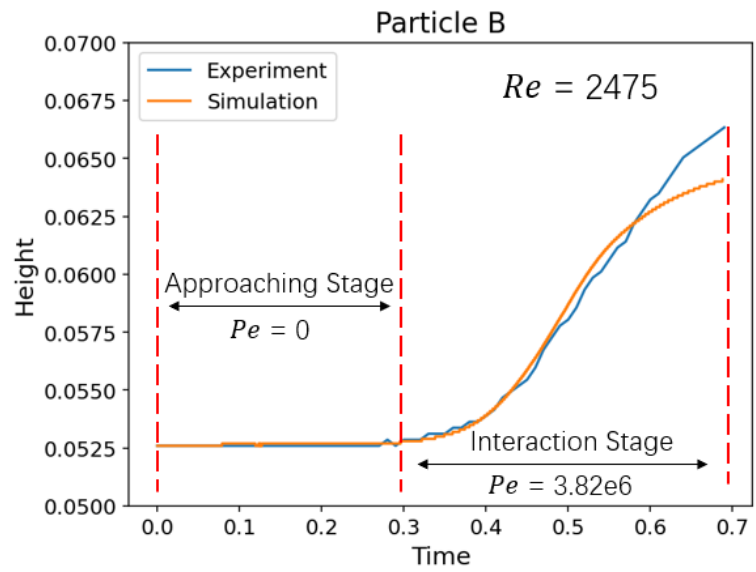

Figure 11: The comparison of the experimental and simulated results on Particle B.

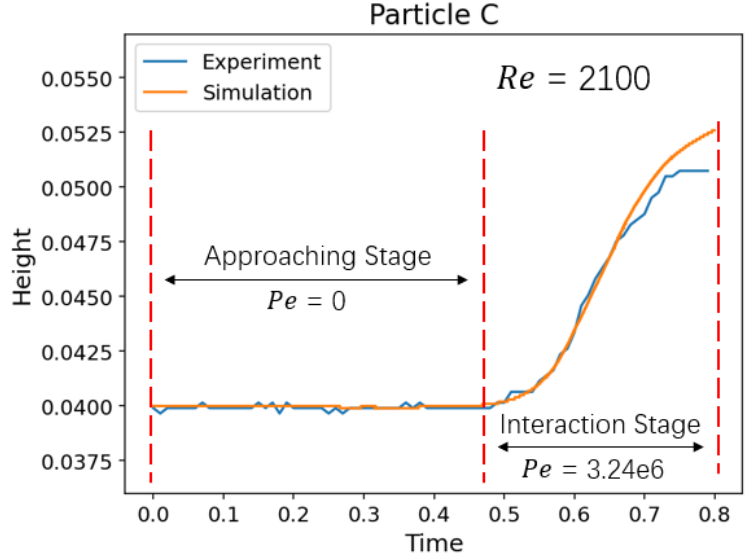

Figure 12: The comparison of the experimental and simulated results on Particle C.

The thermostatic shaking water bath is widely employed in the fields of environmental science, chemical engineering, and bio-engineering. This equipment facilitates the interaction of solute quantities, fluid dynamics, and solid particle behavior, rendering it a representative example of a solute-particle-flow system. Thus, the dispersion with particulate flows in a 3D oscillator is simulated in this section as an application case of the proposed simulation framework. Multiple particles, including particle A - D in Fig. 9, are used. These particles are of different shapes and complexities, providing a diverse set of morphologies for analysis. Additionally, a sphere is also included as the comparison reference. It is worth noting that these particles have been scaled down to the same volume of sphere with radius $R = 5$ lattice units (where 1 unit corresponds to $1 \times 10^{-3}$m, aligning with the LBM space step), while their shape has not been altered. Each simulation will utilize one type of particle with 50 samples. These particles are randomly distributed within a 24R×24R×36R space, with a solute band measuring 24R×24R×1 located at the center of the space (refer to Fig. 13). This solute band consists of 2,000,000 solute particles, each with a mass of $m_s = 0.000125$ g. The diffusion coefficient $D$ of the solutes is set to 2e-9 $m^2/s$. To simulate the oscillation process, a time-dependent acceleration is applied to the DEM particles according to the following expression:

$$\boldsymbol{a}_p = 4\cos\left(\frac{2\pi t}{5000}\right)\boldsymbol{d_i} \tag{37}$$

where $\boldsymbol{d_i}$ is the unit vector for the horizontal direction. This applied acceleration induces a back-and-forth motion of the particles, leading to the spreading of the solute band. And the concentration shares the same definition as stated in Eq. 34.

As for the definition of the dispersion coefficient $D_\alpha$, we follow the one as stated in the work of Derksen [16]. To maintain conciseness within the current context, only the fundamental steps are presented here. Since our simulations consider no gravity and the initial concentration(solute) distribution along the z-axis approximates the point source dispersion, here we only concern the dispersion along the z-axis. Each case of different particles will be carried out multiple times with different initial fluid-particle conditions, to get an averaged concentration distribution along the z-axis. Then, Gaussian distributions are fitted to derive the standard deviations along the time. In the end, a straight line is fitted on the obtained standard deviations to get the scaled coefficient $K$ for the calculation of dispersion coefficient:

$$\frac{C^2}{(2R)^2} = K\frac{U_{\text{slip}}}{2R}t + \text{ offset} \tag{38}$$

where $U_{\text{slip}}$ is defined as:

$$U_{\text{slip}} = \langle u \rangle - \langle u_p \rangle \tag{39}$$

In the work of Derksen[16], $\langle u \rangle$ stands for the velocity in x-direction averaged over the liquid volume and $\langle u_p \rangle$ is the x-velocity averaged over the particle volume, where the time is not related. Here, they are defined as the maximum values along the time for simplification. And this gives the dispersion coefficient $D_\alpha$ as:

$$D_\alpha = \frac{K}{(1-\phi)} \tag{40}$$

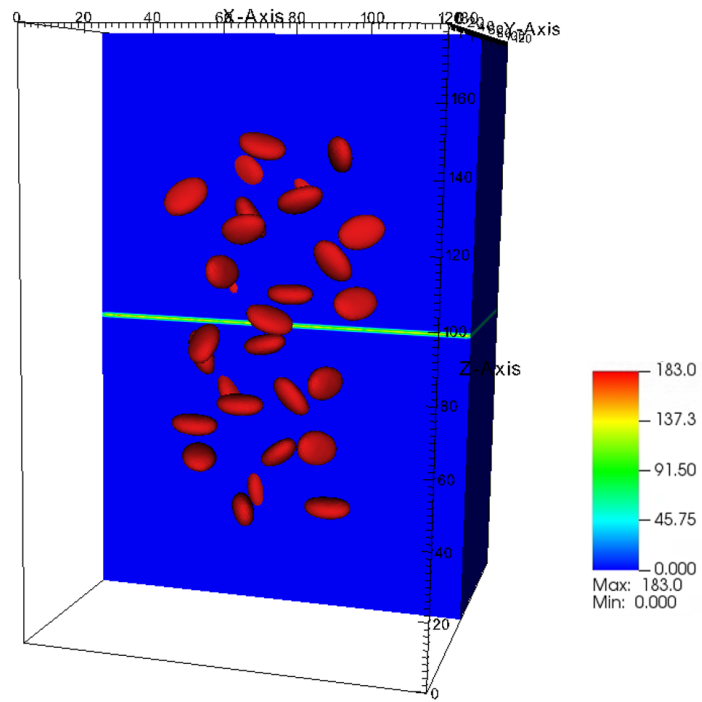

Figure 13: The setup of the application case. This example is shown with the particle D, a realistic cobblestone. And the colors in the background stand for the solute band.

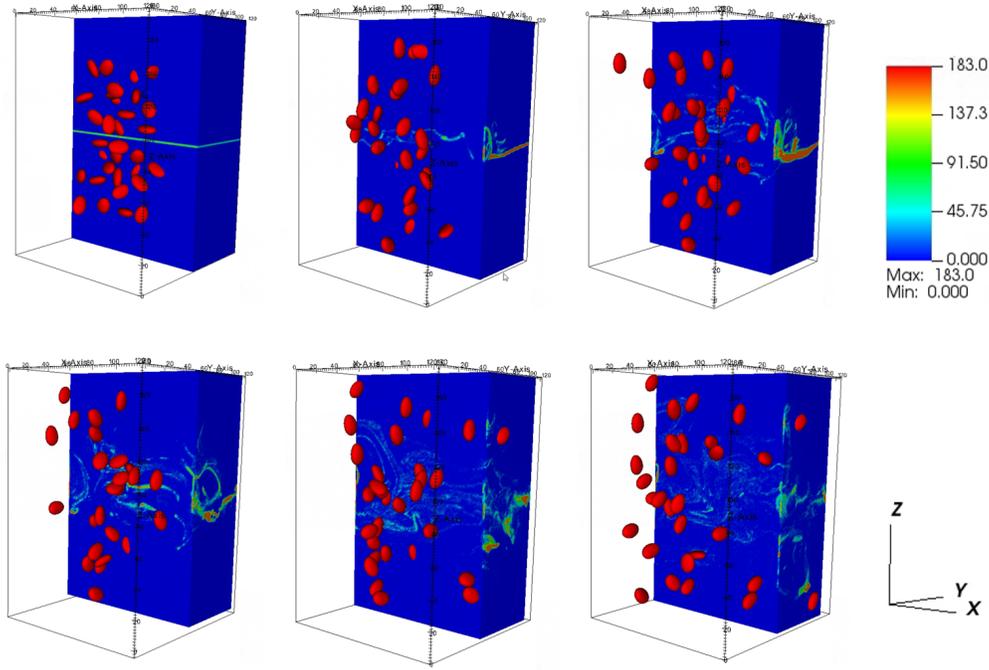

Figure 14: Time series of the standard deviations of the concentration profile of sphere along the z-axis.

Fig. 14 provides a visual representation of the simulation conducted with particle D, illustrating an overview of the simulation process. Fig. 15, 16, 17, 18 and 19 illustrate the simulated results of standard deviations and their fitted lines of the concentration profiles of all involved particles, showcasing their changes over time. The observed time series exhibit similar patterns to those reported in Derksen's work[16], characterized by an initial nonlinear range followed by a linear portion. While it should be noted that the time axis in the current study is significantly longer. Besides, although all five simulations in this study are conducted with the same time steps, the differences in slip velocity $U_{\text{slip}}$ have led to this distinct results in dimensionless time axes. Table 1 presents the calculated dispersion coefficients for each case. The above results demonstrate that variations in particle shape lead to distinct characteristics, including the time required for the system to reach a steady state, the fitted scaled coefficients ($K$), and the calculated dispersion coefficient ($D_\alpha$). These findings highlight the significant impact of particle

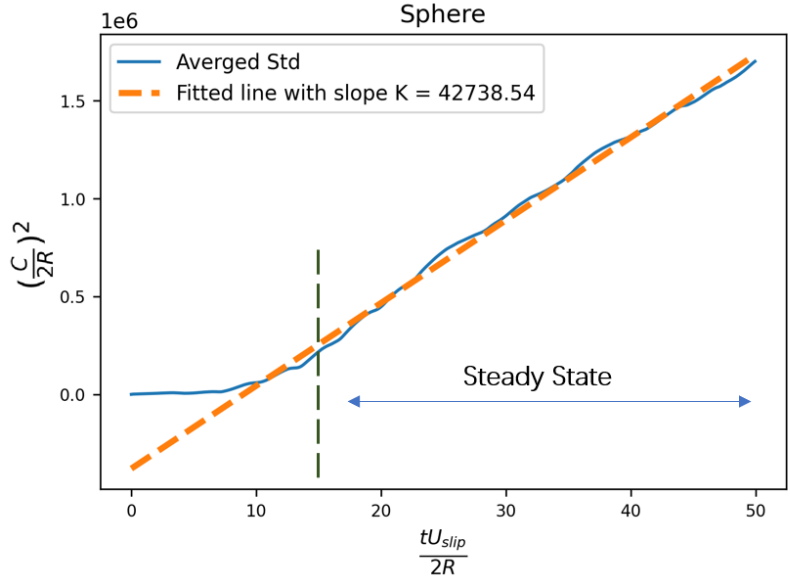

Figure 15: Time series of the standard deviations of the concentration profile of sphere along the z-axis.

morphology on solute transport dynamics in fluid-particle systems.

To get a deeper understanding of the impact of particle shapes on the dispersion coefficients, we introduce several popular particle shape features, including the sphericity $\phi$, circularity $C_c$, nominal diameter $D_n$, surface-equivalent-sphere diameter $D_s$, Corey Shape Factor $CSF$ and the maximum projected area $A_m$.

The sphericity $\phi$ [46] i is a measure of the similarity between a given particle and a sphere, which is defined as:

$$\phi = \frac{A_{ve}}{A} \quad (41)$$

where $A_{ve}$ stands for the surface area of the volume-equivalent sphere to the studied particle and $A$ is the surface area of the studied particle.

The circularity $C_c$ is a frequently used shape measure [47], which evaluates the roundness of non-spherical particles:

$$C_c = \frac{\pi D_s}{P_p} \quad (42)$$

where $P_p$ is the the perimeter of the particle's projected-area.

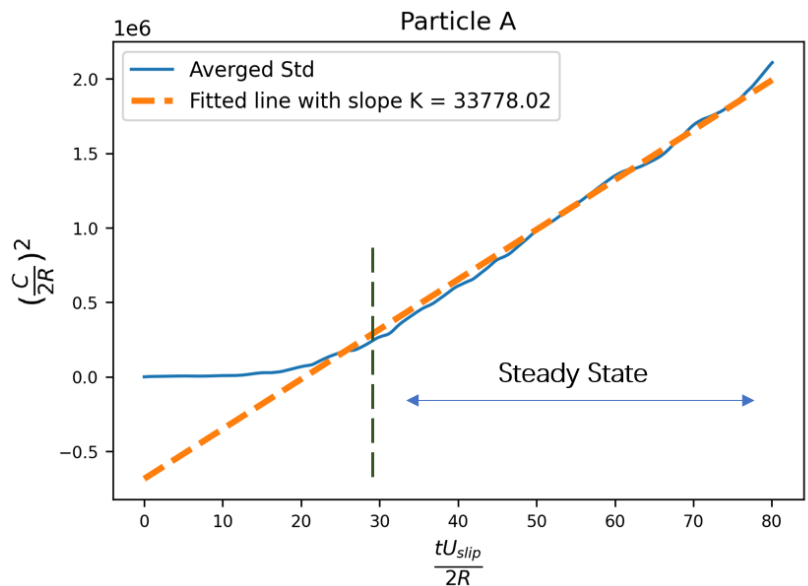

Figure 16: Time series of the standard deviations of the concentration profile of Particle A along the z-axis.

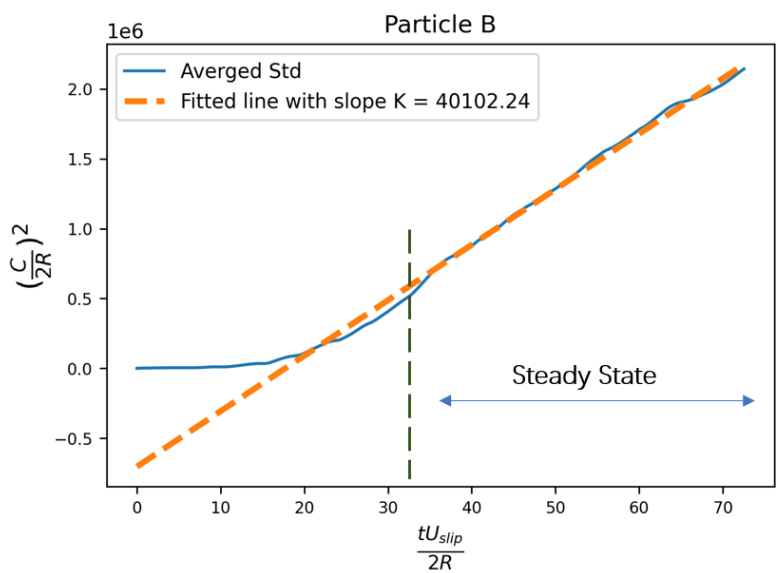

Figure 17: Time series of the standard deviations of the concentration profile of Particle B along the z-axis.

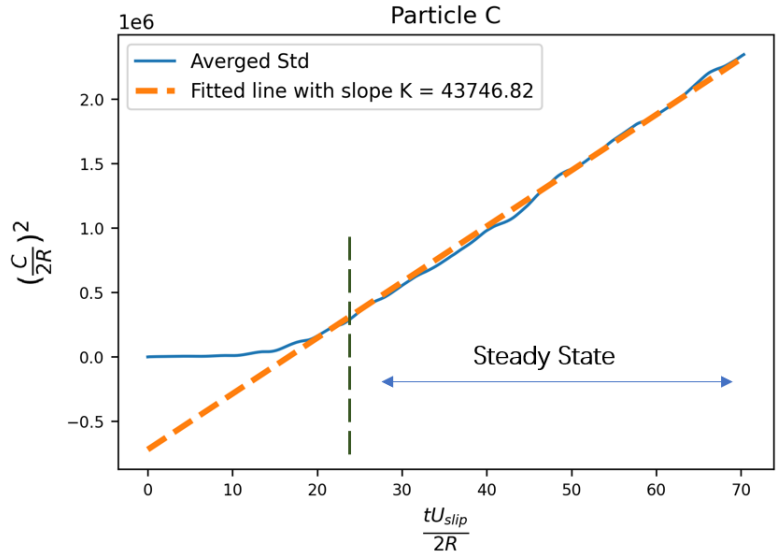

Figure 18: Time series of the standard deviations of the concentration profile of Particle C along the z-axis.

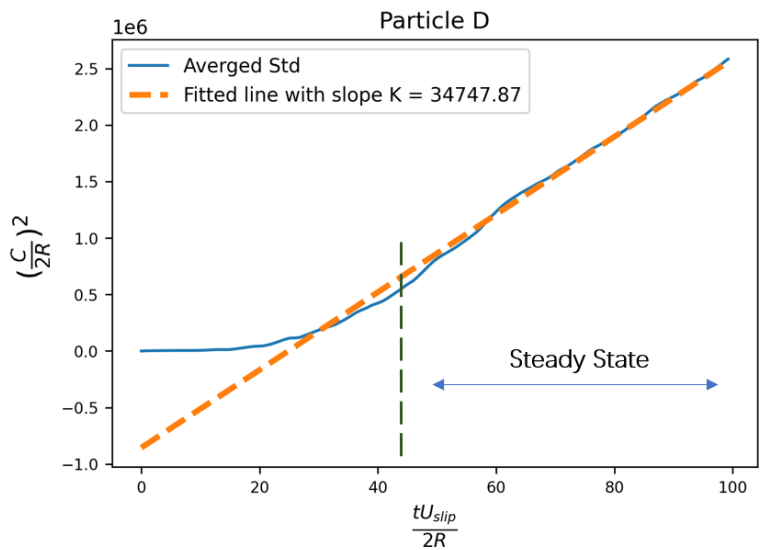

Figure 19: Time series of the standard deviations of the concentration profile of Particle D along the z-axis.

Table 1: The shape features of particles and their corresponding dispersion coefficients

| | $\phi$ | $C_c$ | $D_{ns}$ | $CSF$ | $A_m$ | $D_\alpha$ |
|---|---|---|---|---|---|---|
| Sphere | 1.00 | 1.00 | 1.00 | 1.00 | 78.54 | 42738.54 |
| Particle A | 0.99 | 0.99 | 0.97 | 0.90 | 84.51 | 33778.02 |
| Particle B | 0.97 | 0.98 | 0.92 | 0.73 | 93.28 | 40102.24 |
| Particle C | 0.98 | 0.99 | 0.91 | 0.68 | 95.89 | 43746.82 |
| Particle D | 0.91 | 0.97 | 0.81 | 0.51 | 118.49 | 34747.87 |

The nominal diameter $D_n$ and surface-equivalent-sphere diameter $D_s$ are two popular parameters in the characterization of particle shape [47, 48]. The $D_n$ is defined as the diameter of the volume-equivalent sphere. And the $D_s$ takes the following form:

$$D_s = \sqrt{\frac{4A_p}{\pi}} \quad (43)$$

where $A_p$ = the maximum projected area of the particle. Here, they are combined as $D_{ns} = D_n/D_s$ to form a dimensionless quantity.

Another notable metric is the Corey Shape Factor ($CSF$) [29], which provides insight into the dimensional characteristics of the particles under investigation as given by:

$$\text{CSF} = \frac{L_s}{\sqrt{L_i L_l}} \quad (44)$$

where $L_s$, $L_i$ and $L_l$ are the shortest, intermediate and longest axis lengths of particles.

The calculated shape features of each particle are summarized in Table 1. However, given the limited number of simulation cases, no distinct patterns or correlations between individual shape features and dispersion coefficients can be observed. It is likely that the dispersion coefficient is influenced by a combination of multiple shape features rather than any single parameter in isolation.

Fig. 20 illustrates a heatmap that depicts the relationship between particle shape features and the dispersion coefficient. The heatmap employs the Spearman correlation coefficient [49], which quantifies the monotonic association between variables:

$$r_{spear} = 1 - \frac{6S}{n_o(n_o^2 - 1)} \quad (45)$$

where $n_o$ is the total observation number and $S = \sum d_i^s$, $d_i^s$ refers to the observed rank differences. The Spearman correlation coefficient ranges from -1 to 1 and provides insights into the relationship between variables. A coefficient close to 1 indicates a strong positive relationship, where increasing values of one variable are associated with increasing ranks of the other variable. A coefficient close to -1 indicates a strong negative relationship, where increasing values of one variable are associated with decreasing ranks of the other variable. A coefficient close to 0 suggests little to no monotonic relationship between the variables. This makes the Spearman correlation coefficient a useful tool to detect nonlinear relationships and evaluating different types of monotonic relationships.

As observed from Fig. 20, all shape features exhibit strong correlations among themselves, which suggests a reasonable relationship since each shape feature set corresponds to the same particle. Among these features, $A_m$ displays a distinct negative correlation, whereas the remaining ones demonstrate positive correlations. Furthermore, the analysis emphasizes that variable $C_c$ exhibits the strongest association with the dispersion coefficient, indicating a high level of dependency between them. On the other hand, $CSF$ shows the weakest correlation among the variables under consideration.

To summarize, the present study confirms that particle morphologies significantly influence solute transport in fluid-particle systems. The positive correlations observed between the dispersion coefficients and all shape features, except for $A_m$. Among the shape features, $C_c$ exhibits the strongest correlation with the dispersion coefficient, while $CSF$ demonstrates the weakest association. However, due to the limited number of simulation cases, no clear patterns or direct correlations between individual shape features and dispersion coefficients have been established. The intricate relationship between particle morphology and dispersion coefficients suggests that a comprehensive analysis, incorporating a larger sample size of particles with diverse characteristics, is necessary to fully comprehend the collective impact of particle shapes on solute dispersion in fluid-particle systems. It is possible that the combined effects of various shape features, rather than individual parameters alone, contribute to the observed dispersion coefficients.

## 5. Conclusions

(1) We propose a Random-Walk Metaball-Imaging Discrete Element Lattice Boltzmann Method (RW-MI-DELBM) for 3D solute transport in fluid-

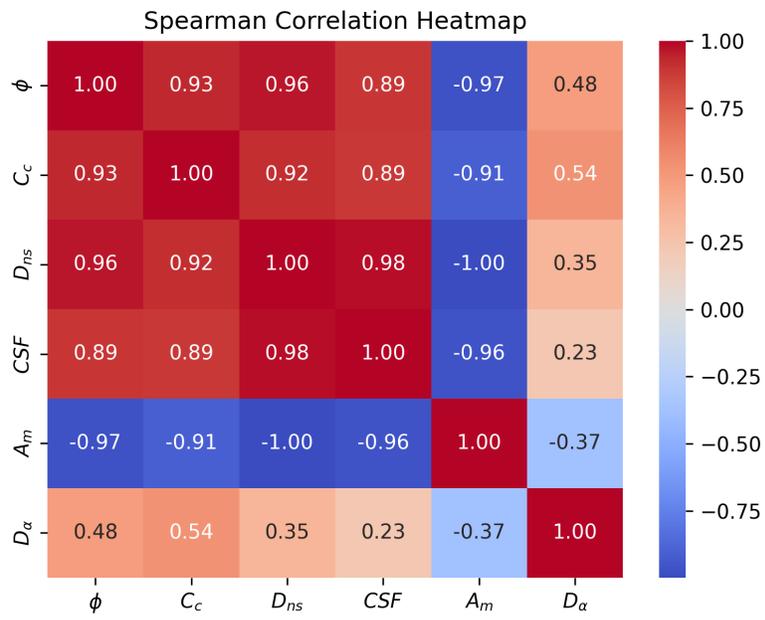

Figure 20: Time series of the standard deviations of the concentration profile with different involved particles along the z-axis.

particle systems with complex granular morphologies.

(2) The proposed framework has undergone thorough validations on various solute transport processes, encompassing both analytical and experimental cases. These validations have successfully demonstrated the effectiveness and accuracy of the framework in simulating the advection and diffusion of solute quantities. Furthermore, the framework has been shown to accurately capture the intricate interactions between solute fields and larger granular materials. The comprehensive validations provide compelling evidence of the framework's efficacy and precision in capturing the complex dynamics of solute transport in fluid-particle systems.

(3) To investigate the impact of particle morphologies on solute transport in fluid-particle systems, we employ the proposed RW-MI-DELBM to simulate solute dispersion in an oscillator. The results clearly indicate that particle morphologies play a significant role in solute transport dynamics.

(4) Among the five selected particle morphologies, namely sphericity ($\phi$), circularity ($C_c$), a dimensionless quantity $D_{ns} = \frac{D_n}{D_s}$ of the nominal diameter $D_n$ and the surface-equivalent-sphere diameter $D_s$, the Corey Shape Factor ($CSF$), and the maximum projected area ($A_m$), all shape features exhibit positive correlations with the dispersion coefficient, except for $A_m$. Notably, $C_c$ exhibits the strongest correlation, while $CSF$ demonstrates the weakest correlation.

(5) The interplay between different particle morphologies and the dispersion coefficients is highly complex. Despite our limited simulations, no clear patterns or direct correlations have been observed. This highlights the necessity for a comprehensive analysis that considers a wider range of shape features and varying conditions to fully comprehend their collective influence on the dispersion coefficient.

## Acknowledgement

We gratefully acknowledge the funds from National Natural Science Foundation of China (Project No.12172305), Natural Science Foundation of Zhejiang Province, China (LHZ21E090002), Key Research and Development Program of Zhejiang Province (Grant No.2021C02048) and Westlake University. We also thank Westlake High-performance Computing Center for computational sources and corresponding assistance. The presented simulations were conducted based on the ComFluSoM open source library.